\documentclass[12pt]{article}

\usepackage{amssymb}
\def\labelmark{}
\def\void{}

{\ifx\void\labelname\def\junk{\end{displaymath}}
\else\def\junk{\end{eqnarray}}\fi\junk\labelmark\def\labelname{}}

\newcommand{\bra}{\begin{array}}
\newcommand{\era}{\end{array}}
\newcommand{\beq}{\begin{equation}}
\newcommand{\eeq}{\end{equation}}
\newcommand{\bqn}{\begin{eqnarray}}
\newcommand{\eqn}{\end{eqnarray}}

\def\BC{\bb C}
\def\_\BC{\bbi C}

\newcommand{\om}{\omega}
 
\newcommand{\la}{\lambda} 
\newcommand{\si}{\sigma}

\newcommand{\ga}{\gamma} 
\newcommand{\te}{\theta}

\newcommand{\pa}{\partial}
\newcommand{\al}{\alpha}  
\newcommand{\ap}{\approx}

\newcommand{\na}{\nabla}
\newcommand{\ka}{\kappa}

\newcommand{\st}{\star}
\newcommand{\ti}{\tilde}
\newcommand{\da}{\dagger}

\newcommand{\lb}{\label}
\newcommand{\ov}{\over}

\newcommand{\ev}{\equiv}
\newcommand{\hb}{\hbar}

\newcommand{\NP}[1]{ {\it Nucl.~Phys.} {\bf #1}}
 
\newcommand{\PL}[1]{ {\it Phys.~Lett.} {\bf #1}}

\newcommand{\PR}[1]{ {\it Phys.~Rev.} {\bf #1}}
\newcommand{\PRL}[1]{ {\it Phys.~Rev.~Lett.} {\bf #1}}

\newcommand{\JP}[1]{ {\it J.~Phys.} {\bf #1}:\  Math.~Gen.~}

\begin{document}
\begin{titlepage}
\renewcommand{\thefootnote}{\fnsymbol{footnote}}

\begin{flushright}
hep-th/0111267
\end{flushright}

\vspace{13mm}
\begin{center}
{\Large\bf Hall Effect in 
Noncommutative Coordinates}

\vspace{17mm}

{\bf{\"{O}mer F. Dayi}$^{a,b}$ 
\footnote{E-mail: {\textsf dayi@itu.edu.tr -- dayi@gursey.gov.tr}}}
\,{and}\,
{ \bf{Ahmed Jellal}$^{a,c}$ 
\footnote{E-mail: {\textsf jellal@gursey.gov.tr -- 
jellal@na.infn.it}}}\\
\vspace{5mm}

\end{center}

\noindent
{\em $^{a}$ {\it Feza G\"{u}rsey Institute, P.O. Box 6, 81220,
\c{C}engelk\"{o}y, Istanbul, Turkey. } }

\vspace{3mm}

\noindent
{\em $^{b}${\it Physics Department, Faculty of Science and
Letters, Istanbul Technical University,\\
80626 Maslak--Istanbul, Turkey.} } 

\vspace{3mm}

\noindent
{\em $^{c}${\it Institute for Theoretical Physics,
University of Stellenbosch, Private Bag X1, Matieland 7601,
South Africa.} }

\vspace{5mm}

\begin{abstract}
We consider electrons in uniform
external  magnetic and electric  fields
which move on a plane whose coordinates are noncommuting. 
Spectrum and eigenfunctions of the related Hamiltonian are obtained.
We derive the electric
 current whose expectation value gives the
Hall effect in terms of an effective magnetic field.
We present a receipt to find the action which can
be utilized in path integrals for noncommuting coordinates.
In terms of this action we calculate the related 
Aharonov--Bohm phase and show that it also yields the same 
effective magnetic field. 
When magnetic field is strong enough this phase becomes independent of 
magnetic field. Measurement of it may give some hints on  spatial noncommutativity.
The noncommutativity parameter $\theta$
can be tuned
such that electrons moving in noncommutative coordinates 
are interpreted
as either leading to  the fractional quantum Hall effect or composite 
fermions in the usual coordinates. 
\end{abstract}

\end{titlepage}

\newpage
\section{Introduction}

To clarify the role which noncommutative coordinates may play in 
physics a better understanding of quantum mechanics in noncommutative 
spaces would be useful. Obviously,
the simplest case is to consider particles
moving in two dimensional noncommutative spaces.
Actually, there exist some realistic physical systems
like
electrons in a  uniform external magnetic field which
are effectively moving in a two dimensional space which is perpendicular to 
magnetic field.
These electrons are investigated in noncommuting coordinates \cite{all}
and interesting phenomena like nonextensive statistics~\cite{dj} and
orbital magnetism~\cite{jel1} are resulted.
We would like to consider electrons moving in two dimensional noncommutative 
space when both uniform external magnetic and electric fields are present.
In the usual case this system leads to Hall effect. Indeed, 
we will show that in noncommuting coordinates one obtains 
Hall effect in terms of an effective magnetic field.

Once noncommutativity is imposed coordinates  behave as operators.
However, we can bring in  noncommutativity
by keeping coordinates as commuting but requiring that composition of their 
functions is given by star product. 
After canonical quantization is performed we
deal with ordinary coordinates but replace ordinary 
product with star product. This procedure leads to an ordinary quantum 
mechanics problem in terms of an effective Hamiltonian 
depending on the noncommutativity parameter $\theta.$
As far as operator description of quantum mechanics is concerned
this procedure suits well.
However, if one deals with path integrals 
the suitable action  should be given in terms of c--number phase space 
variables. One of the possibilities is to find an effective
action which leads to the Green functions which are calculated
in terms of operators \cite{cea}. We will adopt another method:  
The effective action  which  we use in path integrals
is found by replacing  derivatives appearing in
Hamiltonian with c--number momentum
variables. A similar approach is given in \cite{cs} and a different one in \cite{aca}.
We will use the action obtained in this manner in path integrals
to calculate the related Aharonov--Bohm phase
after generalizing the action obtained in symmetric gauge
to embrace the other vector potentials at the first order in $\theta .$
 
In Section 2 we recall how one can find energy eigenvalues and 
eigenfunctions of an electron moving on plane in uniform external 
magnetic and electric fields. This serves as a guide in Section 3
when we deal with the same system in noncommuting coordinates.
In Section 4 we present an approach to derive 
the electric current in noncommutative 
coordinates. Then, we calculate its expectation value utilizing
eigenfunctions derived in Section 3, yielding
the Hall effect in noncommuting coordinates. This can be envisaged as the 
usual Hall effect  
in terms of an effective magnetic field.
Section 5 is devoted to calculate Aharonov--Bohm phase in noncommuting 
coordinates after
presenting our receipt to obtain the action 
suitable to be used in path 
integrals. This phase is used to define  an effective magnetic field
in terms of commuting coordinates.
We observe that effective magnetic fields obtained in Section 4 and in 
Section 5 are the same. 
In the last section
by tuning the parameter $\theta$ and utilizing the effective magnetic field 
we offer two 
different interpretations   of electrons moving in noncommutative space 
as either  leading to the fractional quantum Hall effect~\cite{qhe} or 
composite 
fermions~\cite{cf} in the usual space. Moreover, we propose 
to measure the Aharonov--Bohm phase for
large magnetic fields 
which may give some hints on the existence of spatial
noncommutativity.

\section{Electron Moving on Plane}

An electron moving on the plane $(x,y)$ in
the uniform external electric field
${\vec E}=-{\vec\na}\phi$ and
the uniform external magnetic field $B$  which is perpendicular 
to the plane is described by the Hamiltonian
\beq
\label{fh}
H = {1\over 2m}({\vec p}+{e\over c}{\vec A})^2
-e\phi.
\eeq
We neglect the spin, because taking it
into account  
does not affect our results.

Let us adopt the symmetric gauge
\beq
\label{gco}
{\vec A}=(-{B\over 2}y,{B\over 2}x).
\eeq
During the related experiments the electric field $\vec E$ 
is taken in one of the two possible directions. 
Thus let
the scalar potential be
\beq
\lb{ef}
\phi =-Ex.
\eeq
Making use of (\ref{gco}) and (\ref{ef}) in (\ref{fh})
leads to the Hamiltonian function
\beq
\lb{och}
H(\vec{p}, \vec{r}) =
\frac{1}{2m}\left[
\left( p_x -\frac{eB}{2c} y \right)^2 +
\left( p_y +\frac{eB}{2c} x \right)^2 \right] +eEx .
\eeq
As usual canonical quantization of this system is achieved by 
introducing 
the coordinate and momentum operators $\hat{r}_i,\ \hat{p}_i$ satisfying
\beq
\lb{qun}
[\hat{r}_i,\ \hat{p}_j]=i\hbar \delta_{ij}
\eeq
and dealing with the Hamiltonian operator $\hat{H}$ obtained from
(\ref{och}) as $\hat{H}=H(\hat{\vec{p}},\hat{\vec{r}}).$

To discuss the eigenvalue problem 
\beq
\lb{ep1}
\hat{H} \Psi =E\Psi,
\eeq
it is convenient to 
perform the change of variables 
\[
\hat{z}=\hat{x}+i\hat{y},\qquad  \hat{p}_z =\frac{1}{2} (\hat{p}_x 
-i\hat{p}_y).
\]
and introduce two sets of  
creation
and annihilation operators:
\beq
\lb{ob}
\bra{l}
b^\da =-2i\hat{p}_{\bar z}+{eB\ov 2c}\hat{z}+\la ,\\
b =2i\hat{p}_{z}+{eB\ov 2c}\hat{{\bar z}}+\la ,
\era
\eeq
and 
\beq
\lb{od}
\bra{l}
d=2i\hat{p}_{z}-{eB\ov 2c}\hat{{\bar z}},\\
d^{\da}=-2i\hat{p}_{\bar z}-{eB\ov 2c}\hat{z},
\era
\eeq     
where $\la={mcE\ov B}$. These two sets commute with each other and
satisfy the commutation relations
\beq
\lb{cr}
\bra{l}
[b, b^{\da}]=2m\hb\om,\qquad [d^{\da},d]=2m\hb\om,\\
\era
\eeq
where $\om={eB\over mc}$ 
is the cyclotron frequency. 
Now, the Hamiltonian $\hat{H}$ can be written as
\beq
\label{hc}
\hat{H}= {1\ov 4m}(b^{\da}b+bb^{\da})-
{\la\ov 2m}(d^{\da}+d)-{\la^2\ov 2m}.
\eeq
To calculate the eigenvalues $E$ and the eigenfunctions $\Psi$
we separate
 $(\ref{hc})$  into 
two mutually commuting parts:
\[
\hat{H}=\hat{H}_{osc} -\hat{T} ,
\]
where $\hat{H}_{osc}$
denotes the harmonic oscillator part 
\beq
\lb{ho}
\hat{H}_{osc}= {1\ov 4m}(b^{\da}b+bb^{\da})
\eeq
and the part linear in 
$d$ and $d^{\da}$ is given by 
\beq
\hat{T}={\la\ov 2m}(d^{\da}+d)+{\la^2\ov 2m}.
\eeq

The harmonic oscillator eigenvalue equation 
$\hat{H}_{osc}\Phi_n=E^{osc}_n\Phi_n$ is easily solved:
\beq
\lb{fs}
\bra{l}
\Phi_n= \frac{1}{\sqrt{(2m\hb\om)^n n!}}(b^{\da})^n|0>,\\
E_n^{osc}={\hb\om\ov 2} (2n+1),\qquad n=0,1,2...
\era
\eeq
leading to a discrete spectrum. However, the
eigenvalue equation $\hat{T}\phi=E\phi$
can be analyzed in terms of the eigenvalues of the operators $\hat{r}_i$
denoted by $r_i$ as 
\beq
\lb{fs1}
\bra{l}
\phi_{\al}=e^{i(\al y+i{m\om\ov 2\hb}xy)},\\
E_{\al}={\hb\la\ov m}\al+{\la^2\ov 2m},\qquad \al\in
 \mathbb{R},
\era
\eeq
yielding a continuous spectrum labeled by $\al$.

Therefore, the eigenfunctions and the energy spectrum of the
 Hamiltonian $\hat{H}$ are
\beq
\lb{hfs}
\bra{l}
\Psi_{(n,\al)}=\Phi_n\otimes \phi_{\al}\ev |n,\al>,\\
E_{(n,\al)}={\hb\om\ov 2}(2n+1)-
{\hb\la\ov m}\al-{\la^2\ov 2m}
\qquad n=0,1,2..., \qquad \al\in  \mathbb{R},
\era
\eeq
where $\otimes $ denotes the direct  product.

\section{Electron Moving on Noncommutative Plane}

Let the 
coordinates of the plane be noncommuting: 
\beq
\label{com}
[\mathbf{x},\mathbf{y}]=i\te.
\eeq 
The parameter $\te$ is a real constant. 
Noncommutativity can be imposed 
by treating the coordinates as
commuting but requiring that composition of their functions 
is given in terms of 
the star product 
\beq
\label{star}
\st\equiv\exp \frac{i\te}{2} \Big(
{\stackrel\leftarrow\pa}_{x} {\stackrel\rightarrow\pa}_{y}
-{\stackrel\leftarrow\pa}_{y}{\stackrel\rightarrow\pa}_{x}
\Big).
\eeq 
Now, we deal with  the commutative coordinates $x$ and $y$ but
replace the ordinary products  with the star product 
(\ref{star}). For example, 
instead of the commutator (\ref{com}) one defines 
\beq
x\st y-y\st x=i\te .
\eeq 

We would like to study the Hamiltonian (\ref{och})
in terms of the noncommutative coordinates (\ref{com}).
First we quantize  this system 
by establishing the commutation relations (\ref{qun}).
Then, the noncommutativity of the coordinates  is taken into account
by defining a new operator as 
\beq
\lb{dnc}
{\hat H}\st \Psi (\vec{r})
\equiv {\hat H}_{nc} \Psi (\vec{r}) .
\eeq
This definition yields the Hamiltonian operator
$\hat{H}_{nc}$ 
\beq
\lb{nh}
{\hat H}_{nc} =
\frac{1}{2m}\left[
\left( (1-\kappa )\hat{p}_x -\frac{eB}{2c} \hat{y} \right)^2 +
\left( (1-\kappa )\hat{p}_y +\frac{eB}{2c} \hat{x} \right)^2 \right]
+eE(\hat{x}-{\te\ov 2\hb}\hat{p}_y),
\eeq
when the coordinate representation of momentum $\hat{p}_i=-i\hbar \partial_i$
is used and  $\kappa =\frac{e\theta B}{4\hbar c}.$

The eigenvalue problem  
\beq
\lb{evp2}
\hat{H}_{nc} \Psi^{nc} =E^{nc}\Psi^{nc} 
\eeq
is as in ordinary quantum mechanics
in spite of the fact that electron moves on noncommutative plane.
The solutions of this problem can be worked out in a manner similar to the 
one used in the previous section
although here there exists a term linear in momentum which was not present 
in $\hat{H}$.
Then, let us introduce
two sets of  operators
\beq   
\lb{nob}
\bra{l}
{\ti b}^{\da}=-2i\hat{{\ti p}}_{\bar z}+{eB\ov 2c}\hat{z}+\la_-, \\
{\ti b}=2i\hat{{\ti p}}_{z}+{eB\ov 2c}\hat{\bar z}+\la_-,
\era
\eeq
and 
\beq
\lb{nod}   
\bra{l}
{\ti d}=2i\hat{{\ti p}}_{z}-{eB\ov 2c}\hat{{\bar z}},\\
{\ti d}^{\da}=-2i\hat{{\ti p}}_{\bar z}-{eB\ov 2c}\hat{z},
\era
\eeq
where $\hat{{\ti p}}_{z}=\ga \hat{p}_z$; 
$\ga=1-\ka$. The real parameter $\la_-$ will be fixed later.
The sets of operators $({\ti b},{\ti b}^\da )$ and
$({\ti d},{\ti d}^\da )$ commute with each other. Moreover,
they satisfy the
commutation relations 
\beq   
\lb{ncr}
\bra{l}
[{\ti b} ,{\ti b}^{\da}]=2m\hb{\ti\om},\qquad 
[{\ti d}^{\da},{\ti d}]=2m\hb{\ti\om},\\
\era
\eeq
where ${\ti\om}=\ga\om$.
The Hamiltonian $\hat{H}_{nc}$ can be written as
\beq
\label{nh1}
\hat{H}_{nc}= {1\ov 4m}({\ti b}^{\da}{\ti b}+
{\ti b}{\ti b}^{\da})-{\la_+\ov 2m}
({\ti d}^{\da}+{\ti d})-{\la_-^2\ov 2m},
\eeq
where the parameters  $\la_{\pm}$ are fixed to be
\beq
\lb{lpn}
\la_{\pm}=\la\pm{emE\te\ov 4\ga\hb}.
\eeq

We take into account only 
the values of the noncommutativity 
parameter 
$\theta\ne 4\hbar c/eB.$
Otherwise,   $\la_\pm$ diverge. We will give a brief discussion of
this fact in the last section.

Observing the similarity between the Hamiltonians (\ref{hc})
and (\ref{nh1}) the solutions of the eigenvalue problem (\ref{evp2})
can be read from  (\ref{hfs}) as
\beq   
\lb{nfs}   
\bra{l}
\Psi^{nc}_{(n,\al,\te)}\ev|n,\al,\te>=
{1\ov \sqrt{(2m\hb{\ti \om})^n n!}}
e^{i(\al y+{m\ti{\om}\ov 2\hb}xy)}({\ti b}^{\da})^n|0>,\\
E^{nc}_{(n,\al,\te)}=
{\hb{\ti\om}\ov 2}(2n+1)-
{\hb\ga\la_+\ov m}\al-{m\ov 2}\la_-^2 
\qquad n=0,1,2...,\qquad \al\in \mathbb{R}.
\era
\eeq
We would like to emphasize  that the results of the previous section are 
recovered if the noncommutativity parameter
$\te$ is switched off.

\section{Hall Conductivity on Noncommutative Plane}

We would like to find conductivity resulting from 
the Hamiltonian $\hat{H}_{nc}.$ The first step in this direction
is to define the related current. Although the identification of 
derivatives with 
momentum operators $\partial_i={i \ov \hbar } \hat{p}_i$ is only valid in
coordinate representation, we will use  this definition in defining the
current operator $\hat{\vec{J}}$ on noncommutative plane as
\beq
\lb{nco}
{\hat{\vec{J}}} ={ie\rho \ov \hbar } [ \hat{H}_{nc},  \hat{\vec{r}} \  ]
=
{e\ga \rho\ov m} ( \gamma  \hat{\vec p }+{e\ov c}{\vec A}+{\vec a} ),
\eeq
where ${\vec a}=(0,-{meE\te\ov 2\hb\ga})$ and $\rho$ denotes electron 
density.

Now, the expectation value of the current operator 
 $<\hat{\vec{J}}>$ can be calculated 
with respect to the eigenstates $|n,\al,\te>$
 (\ref{nfs}) leading to
\beq
\lb{ncco}
\bra{l}
< \hat{J}_x >=0,\\
<\hat{J}_y>=-\ga\Big({\rho ec\ov B}\Big)E.
\era
\eeq
Therefore, the Hall conductivity on noncommutative plane,
denoted by $\si_H^{nc}$, is
\beq
\lb{nhc}
\si_H^{nc}=-\ga\Big({\rho ec\ov B}\Big).
\eeq
Recall that
in the ordinary case the Hall conductivity 
$\sigma_H$ and the filling factor $\nu$ are given as
\beq
\lb{ohc}
\si_H={ e^2\ov h}\nu ,\qquad  \nu={\Phi_0 \rho \ov B },
\eeq
where $\Phi_0 = hc/e.$ 
Comparison of
(\ref{nhc}) with (\ref{hc}) suggests that one can interpret the
noncommutative case as a theory of Hall effect on commuting plane 
 with an effective magnetic field
\beq   
\lb{nmf}
B_{eff}={B\ov 1-{e\te B\ov 4 \hbar c}}.
\eeq
Moreover, 
the effective filling factor
\beq
\lb{rff}
\nu_{eff}={\Phi_0 \rho \ov B }(1-{e\te B\ov 4 \hbar c}),
\eeq
can also be defined.

\section{The Aharonov--Bohm Effect}
We would like to calculate the Aharonov--Bohm effect 
on noncommutative plane
by examining the action appearing in the related path 
integral. When we deal with  quantum mechanics in 
the usual spaces it
is the related classical action.
However, it is not clear what should be the definition 
of action appropriate for path integrals
when  noncommutativity
is taken into account.
Because, we define  Hamiltonian operators 
in terms of the receipt 
used in (\ref{dnc})  where,
we identify $\partial_i \equiv (i/\hbar )\hat{p}_i.$
We propose to 
define the path integral in noncommutative space as
\beq
\lb{pf}
Z=\int d^2 p \;d^2 r\; 
e^{\frac{i}{\hbar} \int dt\;[{\vec p}\cdot \dot{\vec r}-H_{\rm eff} 
({\vec r},{\vec p})]},
\eeq
where $({\vec r},\ {\vec p})$ define the commuting  phase space
and $H_{ {\rm eff } }({\vec r},{\vec p})$ will be obtained from
the related Hamiltonian operator in noncommutative space  by 
replacing 
the operators $\hat{\vec{p}},\ \hat{\vec{r}}$ with  c--number
variables $\vec{p},\ \vec{r}.$

Let us deal with the  Hamiltonian operator on  noncommutive plane 
in the constant external electric field  ${\vec E}=(E_x,E_y)$
and the constant magnetic field $B$ in the symmetric gauge (\ref{gco}):
\beq
\lb{fef}
\hat{H}^\prime_{nc} =
\frac{1}{2m}\left[
\left( \gamma \hat{p}_x -\frac{eB}{2c} \hat{y} \right)^2 +
\left( \gamma \hat{p}_y +\frac{eB}{2c} \hat{x} \right)^2 \right]
+eE_x(\hat{x}-{\te\ov 2\hb}\hat{p}_y) 
+eE_y(\hat{y}+{\te\ov 2\hb}\hat{p}_x).
\eeq
Although, the Hamiltonian operator
\beq
\lb{h10}
\hat{H}_{\theta }\equiv {\ga^2\ov 2m}{\hat {\vec p\ }}^2+
{e^2\ov 2mc^2}{\vec A}^2 +
{e\ga\ov 2 mc } ({\hat {\vec p}} \cdot {\vec A} +
{\vec A} \cdot {\hat {\vec p}} ) +
{\hat {\vec p}} \cdot {\vec K} +e\vec{E}\cdot \hat{\vec{r}} ,
\eeq
where ${\vec K}={e\te\ov 2\hb}(-E_y,E_x),$
is equivalent to (\ref{fef}) only when the vector potential $\vec{A}$
is as given in (\ref{gco}), we assume that at least at the first 
order in $\theta$ it is valid for any gauge potential.

The c--number effective Hamiltonian corresponding to (\ref{h10}) is
\beq
\lb{efnc1}
H_{\rm eff}={\ga^2\ov 2m}{\vec p\ }^2+
{e^2\ov 2mc^2}{\vec A}^2 +
{\vec p}\cdot \left({e\ga\ov  mc } {\vec A}
+{\vec K}\right ) 
+e\vec{E}\cdot \vec{r} .
\eeq
Thus, the  partition function
can be written as
\beq
\lb{pf1}
Z_{nc}=N\int d^2 p \;d^2 r\;
e^{\frac{i}{\hbar}\int_{t_1}^{t_2} dt\;[{\vec p}\cdot (\dot{\vec r}-
{e\ga\ov mc }{\vec 
A}-{\vec K})-{\ga^2\ov 2m}{\vec p}^2-
{e^2\ov 2mc^2}{\vec A}^2-e\vec{E}\cdot \vec{r}  ]},
\eeq
where $N$ is a normalization constant.
Now, we can integrate over the momenta $\vec{p}$ to obtain
\beq
\lb{pf2}
Z_{nc}=\frac{2m}{\ga^2} N \int d^2r \;
e^{\frac{i}{\hbar}\int_{t_1}^{t_2} dt\;[{m\ov 2\ga^2}\dot{\vec r}^2
-{m\ov\ga^2}\dot{\vec r}\cdot({e\ga\ov mc}{\vec A}+{\vec K})+
{e\ov c\ga}{\vec A}\cdot {\vec K}+{m\ov 2\ga^2}{\vec K}^2
-e\vec{E}\cdot \vec{r} ]}.
\eeq
Because of being a constant
$K^2$ term is irrelevant.
  In terms of a new
normalization constant $N^\prime$ we can write 
\beq
\lb{spi}
Z_{nc}= N^\prime \int d^2r \;
e^{\frac{i}{\hbar}S_0 
-{im\ov \hbar \ga^2}
\int_{t_1}^{t_2} dt\;
\dot{\vec r}\cdot({e\ga\ov mc}{\vec A}+{\vec K}) } ,
\eeq
where we defined
\beq
\lb{s0}
S_0=\int_{t_1}^{t_2} dt\ \left[ {m\ov 2\ga^2}\dot{\vec r}^2
-e\vec{E}\cdot \vec{r} 
+{e\ov c\ga}{\vec A}\cdot {\vec K} \right] .
\eeq

The last term in the exponent of (\ref{spi}) can be 
written as
\beq
\lb{np}
i\delta = -{im\ov \ga^2\hbar} \int_{\vec{r}(t_1)}^{\vec{r}(t_2)} 
d \vec{r} \cdot ({e\ga\ov mc} {\vec A}+{\vec K} ).
\eeq

To investigate  the Aharonov--Bohm effect 
for noncommuting coordinates, let
$\vec{A}=\vec{\nabla} f(\vec{r}).$ 
Then
\beq
\lb{phas}
\delta =
-{m\ov \ga^2\hbar} \int_{\vec{r}(t_1)}^{\vec{r}(t_2)} 
 d{\vec r}\cdot ({e\ga\ov mc}{\vec \nabla }f(\vec{r})+{\vec K})
\eeq
depends only on the points 
${\vec{r}(t_1)},\ {\vec{r}(t_2)} $ which are kept fixed in path integrals.
Therefore, it is a phase factor. i.e. propagation with the action $S_0$
is changed up to the phase factor (\ref{phas}).

The Aharonov--Bohm effect can now be calculated as the integral of  the 
phase factor (\ref{phas}) along a loop enclosing a magnetic flux.
${\vec K}$ is a constant vector so that, it does not 
contribute
to the Aharonov--Bohm 
phase:
\beq
\lb{ka}
\oint d{\vec r}\cdot {\vec K}=0.
\eeq
As in the ordinary case the unique contribution  
is due to the gauge 
potential
\beq
\lb{np2}
\Phi_{AB}^{nc}=-2\pi {BS\ov \ga \Phi_0 },
\eeq
where $S$ denotes the surface enclosed.
Obviously, when $\theta =0$ the usual
Aharonov--Bohm phase results
\beq
\lb{npo}
\Phi_{AB}=-2\pi {BS\ov  \Phi_0 }.
\eeq
Thus we can envisage the noncommutative case as a 
theory in commuting coordinates with an effective magnetic field
\beq   
\lb{nmf2}
B_{eff}={B\ov 1-{e\te B\ov 4 \hbar c}},
\eeq
which is the one obtained previously (\ref{nmf}).

\section{Discussions}

Electrons moving on a noncommutative plane when
uniform external magnetic and electric fields are present 
can be envisaged as the usual motion of electrons 
experiencing an effective magnetic field (\ref{nmf}). This is one of 
our  main results. It followed by considering either Hall effect or 
Aharonov--Bohm phase in noncommuting coordinates. 
By tuning
the value of the \mbox{noncommutativity} parameter $\theta$ we can offer
two different interpretations of this fact:

\noindent
{\bf The  fractional quantum Hall effect} is one of the most interesting
features of low dimensional systems~\cite{qhe}. For electrons moving on a 
plane  in a magnetic field which is perpendicular to the plane and 
a uniform external electric field which is in the plane, the observed 
Hall conductivity is 
\[
\si_H=f \;{e^2\ov h},
\] 
where $f=1/3,\ 2/3,\  1/5,\ \cdots ,$ denoting the fractional quantized values of the 
filling factor $\nu .$
We would like to interpret this phenomena, which is known as the fractional 
quantum Hall effect, in terms of the Hall effect on noncommutative plane.
More precisely we identify the  
effective filling factor (\ref{rff}) with the observed value $f$ by fixing 
the value of $\theta$ to be $\theta_H:$ 
\beq
\lb{fff}
\nu_{eff}|_{\te=\te_H}=f .
\eeq
In fact, this can be solved as
\beq
\lb{fff1}
{\te_H}={2\Phi_0\ov\pi B}\Big(1-f{B\ov\Phi_0\rho}\Big).
\eeq
Therefore, when 
$\te$  is fixed to be $\theta_H$
one can envisage the Hall effect on
noncommutative plane as the usual fractional quantum Hall effect.

\noindent
{\bf Composite fermions} are new kind of particles appeared in 
condensed 
matter physics to provide an explanation of the behavior of electrons 
moving on plane when a strong magnetic field $B$ is present~\cite{cf}. 
Electrons possessing $2p$; $p=1,2,\cdots$, flux quanta (vortices) can be 
thought of being composite fermions. One of the most important features of 
them is they feel effectively the magnetic field
\beq
\lb{cfm}
B^*=B - 2p\Phi_0\rho,
\eeq
where $\rho$
is the electron density.
To interpret electrons moving on noncommutative space in the magnetic 
field $B$ 
as the usual composite fermions we should tune $\theta$ such that
\beq
\lb{emf}
B_{eff}|_{\te=\te_c}=B^{*},
\eeq
where $B_{eff}$ is given in (\ref{nmf}).
We solve this to obtain
\beq
\lb{cte}
\te_c={2\Phi_0\ov\pi B}\Big[1-(1-2p{\rho\Phi_0\ov B})^{-1}\Big],
\eeq
which in the limit of strong magnetic field leads to
\beq
\lb{cte1}
\te_c\ap 4p{\rho\ov\pi} \Big({\Phi_0\ov B}\Big)^2.
\eeq
Thus, composite fermions  can be envisaged
as electrons moving in noncommutative plane in the magnetic field $B$ and
the electric field $E$ when we fix
$\te =\te_c$. 

{\bf The Aharonov--Bohm phase }$\Phi_{AB}^{nc}$ possesses
a very interesting limit.  
Let us deal with $ \theta \neq 0$ and
the  magnetic field satisfying the condition 
\beq
\lb{lab}
B\gg {4c\hbar \ov e\theta} .
\eeq
For these values of magnetic field the Aharonov--Bohm phase
$\Phi_{AB}^{nc}$ (\ref{np2}) becomes
\beq
\lb{np3}
\Phi_{AB}^{nc} \approx {4c\hbar S \ov e\theta \Phi_0} ,
\eeq
which  is independent of $B$. For $\vec{E}=0$ the action $S_0$ 
reads
\[
\tilde{S}_0\equiv S_0|_{\vec{E}=0}=
\frac{m}{2\gamma^2}\int dt \ {\dot{\vec{r}}}^2,
\]
which depends on the magnetic field as $B^{-2}$ when
(\ref{lab}) is satisfied. 
If the phase $\Phi_{AB}^{nc}$
can be measured for the particle propagating with the action 
$\tilde{S}_0$ 
when $B$ satisfies (\ref{lab}) and observed that it is independent of $B$ 
after a certain value of $B,$
it may be an evidence for spatial noncommutativity.
Obviously, this conclusion is valid only for small values
of $\theta .$ Because we assume it when  we write the action (\ref{h10}).  
In \cite{glr}  Aharonov--Bohm effect in 
noncommutative coordinates was studied in terms of a field theoretical 
approach where an 
 experiment to detect spatial noncomutativity was proposed.

{\bf Critical value} of $\theta$ defined as 
\beq
\lb{cp}
\theta^* = {4\hbar c \ov eB} 
\eeq
is avoided through this work.
At this value of $\theta$  all of our  analysis fail,
because $\gamma =1- eB\theta / 4\hbar c $ cannot be inverted,
thus $\la_\pm $ (\ref{lpn}) are not well defined. 
Indeed, when $\theta=\theta^*$
the Hamiltonian in  noncommutative coordinates $(\ref{nh})$ becomes 
\beq
\lb{nh2}
\hat{H}_{nc}(\te=\te^*) ={m\om^2\ov 8}({\hat x}^2+{\hat y}^2)+eE({\hat x}-
{\te^* \ov 2\hb}{\hat p}_y).
\eeq
i.e. the terms quadratic in momenta disappear, however a term linear in 
momenta survives.  Obviously, this system should be studied separately.

\begin{center}
\vspace{1cm}
\noindent
{\bf ACKNOWLEDGEMENTS}
\end{center}

We would like to thank I. H. DURU for fruitful discussions on the Aharonov--Bohm effect.

\pagebreak

\end{document}